\documentclass{epl}
\usepackage{graphicx}

\newcommand{\idest}{{\it i.e.\/}}
\newcommand{\Journal}[4]{#1 {\bf #2}, #3 (#4)}

\newcommand{\oneu}{$1_u$}
\newcommand{\twou}{$2_u$}
\newcommand{\zeroup}{$0_u^+$}
\newcommand{\singS}{$1^1$S}
\newcommand{\tripS}{$2^3$S$_1$}
\newcommand{\tripP}{$2^3$P$_2$}
\newcommand{\quintet}{$^5\Sigma_g^+$}
\newcommand{\singlet}{$^1\Sigma_u^+$}
\newcommand{\ee}[1]{10$^#1$}

\title{Rotationally induced Penning ionization of ultracold photoassociated helium dimers}

\shorttitle{Rotationally induced Penning ionization...}

\author{J. L\'eonard \inst{1} \thanks{Permanent address: Institut de Physique et Chimie
des Mat\'eriaux de Strasbourg, France.} \and A. P. Mosk \inst{1}
\thanks{Permanent address: Dept. of Science \& Technology and
MESA+ research institute, University of Twente, The Netherlands.}
\and M. Walhout \inst{1}\thanks{Permanent address: Calvin College,
Grand Rapids, MI, USA.} \and M. Leduc
\inst{1} \and \\
M. van Rijnbach\inst{2} \and D. Nehari\inst{2}
\and P. van der Straten\inst{2}}

\shortauthor{J. L\'eonard \etal}

\institute{ \inst{1}Ecole Normale Sup\'erieure and Coll\`ege de
France, Laboratoire Kastler Brossel -\\24 rue Lhomond, 75231 Paris
Cedex 05,
France.\\
\inst{2} Debye Institute, Atom Optics and Ultrafast Dynamics,
Utrecht University -\\P.O. Box 80,000, 3508 TA Utrecht, The
Netherlands.}




\begin{document}

\maketitle

\begin{abstract}

We have studied photoassociation of metastable \tripS\ helium
atoms near the \tripS-\tripP asym\-ptote by both ion detection in
a magneto-optical trap and trap-loss measurements in a magnetic
trap. A detailed comparison between the results of the two
experiments gives insight into the mechanism of the Penning
ionization process. We have identified four series of resonances
corresponding to vibrational molecular levels belonging to
different rotational states in two potentials. The corresponding
spin states become quasi-purely quintet at small interatomic
distance, and Penning ionization is inhibited by spin conservation
rules. Only a weak rotational coupling is responsible for the
contamination by singlet spin states leading to a detectable ion
signal. However, for one of these series Bose statistics does not
enable the rotational coupling and the series detected through
trap-loss does not give rise to sufficient ionization for
detection.

\end{abstract}


Recently there have been many experimental efforts to achieve
Bose-Einstein condensation (BEC) for metastable rare-gas atoms
~\cite{Orsay,ENS,VU,Eindhoven,Hannovre}. The achievement of BEC
with ultracold metastable $^4$He(\tripS) atoms
(He*)~\cite{Orsay,ENS} followed the
prediction~\cite{Shearer,Muller} that spin conservation prohibits
Penning ionization in the fully stretched molecular spin state and
thus prevents the cold cloud of spin-polarized metastable atoms to
ionize before it Bose condenses. By contrast, Penning ionization
is less effectively suppressed in heavier rare gases, which must
be trapped in a metastable $^3$P state and are therefore subject
to stronger spin-orbit collisional couplings~\cite{Doery}. In this
context, there is general interest in both experimental and
theoretical studies of the dynamics that lead to Penning
ionization and its suppression in trapped metastable rare gases.
For a magnetically trapped sample in the He* state, the common
spin orientation imposes a strict conservation rule that permits
collisional Penning ionization (He*(\tripS) + He*(\tripS)
$\rightarrow$ He[\singS] + He$^+$ + e$^-$) only through a
spin-dipole coupling that is relatively weak~\cite{Shlyapnikov}.
The goal of this paper is to discuss the mechanism leading to
Penning ionization that we have observed in our photoassociation
experiments with helium. Through a detailed comparison between
data from two qualitatively different experiments, we show that
weak couplings induced by molecular rotation can lead to the
ionization of photoassociated molecules.

\begin{figure}
\twofigures[scale=0.67]{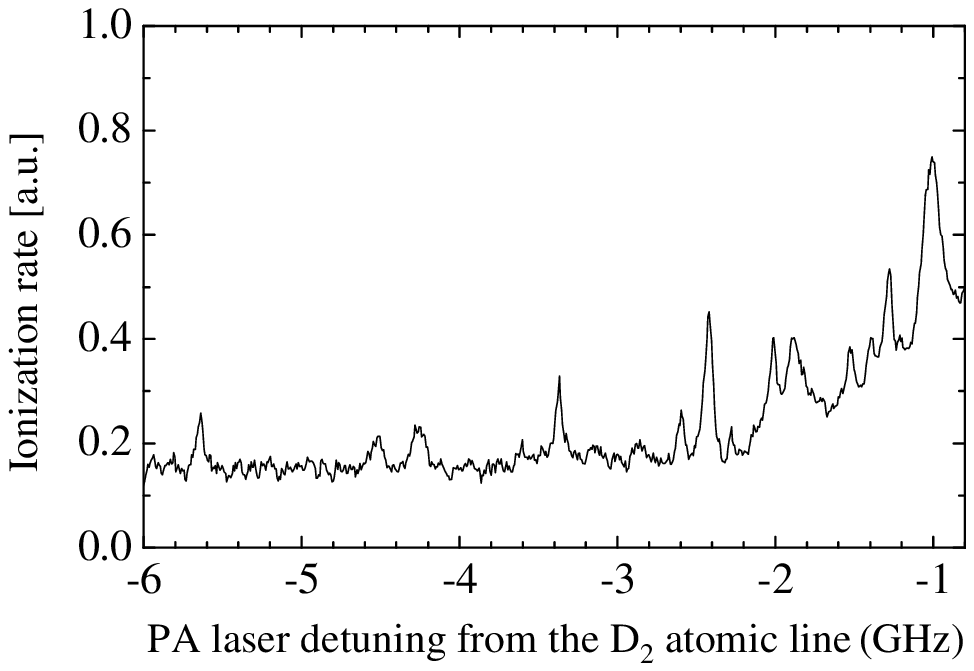}{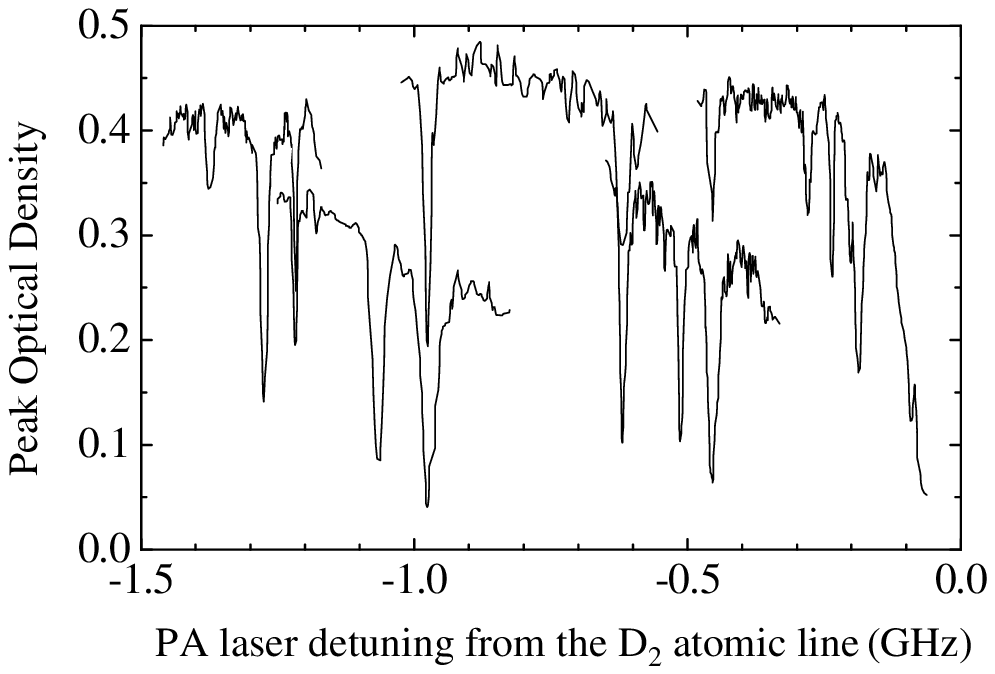} \caption{Typical
data from the Utrecht Magneto-Optical Trap (MOT) experiment:
ionization rate versus detuning of the PA laser from the D$_2$
atomic line. This spectrum is the result of the average over 20
scans. The PA laser intensity is \ee{4}~$I_{sat}$.
\label{Fig:Data_Utrecht}} \caption{Typical data from the ENS
magnetic trap experiment: optical density versus detuning of the
PA laser from the D$_2$ atomic line. The drops in peak optical
density result both from trap loss and temperature increase of the
sample. The five spectra displayed are obtained with different PA
laser exposure times (from 30 ms to 150 ms) and intensities (from
$3I_{sat}$ to $I_{sat}/10$). Each spectrum is a series of up to
500 individual data points smoothed by averaging over 5 adjacent
points. \label{Fig:Data_ENS}}
\end{figure}

We have performed complementary photoassociation (PA) experiments close
to the \tripS-\tripP\ asymptote in Utrecht~\cite{Utrecht,Hersbach} and
Paris~\cite{ENS2}. In the experiments we observe a molecular spectrum of
photoassociation resonances up to 10 GHz below the asymptote. Our ability
to interpret the molecular spectrum depends critically on the fact that
we use different methods of trapping and detection in the two
experiments. In Utrecht a MOT operated at a wavelength of 1.083 $\mu$m is
used to accumulate typically a few $10^6$ He* atoms at a temperature of
$\sim$ 1.9 mK and a density of order $10^{10}$ cm$^{-3}$. The
accumulation of the PA spectrum is done by pulsing the frequency of the
MOT beams from the optimized trapping frequency to far off-resonance
(about 200 natural linewidths) at the rate of 25 kHz with a duty cycle of
50 \%  and shining the PA laser during the off-resonant periods. When
scanning the PA laser frequency, PA resonances are detected as peaks in
the ion production rate, which is measured with a multichannel plate
(MCP). Thus, this experiment is extremely sensitive to those excited
states that predominantly decay by Penning ionization. Fig.
\ref{Fig:Data_Utrecht} shows a typical spectrum obtained after averaging
over 20 scans. The high rate at which the MOT and PA lasers are pulsed
allows for a fast accumulation of the PA spectrum over broad frequency
ranges (from 0 to 20 GHz below the D$_2$ line). Since the clouds are
unpolarized, the PA laser can excite molecular states corresponding to
both {\em gerade} and {\em ungerade} symmetries.

In Paris typically 5$\times 10^8$ atoms are loaded in a magnetic trap
from a MOT operating at 1.083 $\mu$m. Forced evaporative cooling is used
to bring the spin-polarized cloud to densities of $10^{13}$ cm$^{-3}$ and
temperatures of typically $10\,\mu$K or lower, that is to say slightly
above the critical temperature for BEC. The cloud containing a few $10^6$
atoms is then illuminated by a PA light pulse, released from the trap and
destructively imaged to determine its density and temperature. PA
resonances induce loss in the magnetic trap, which can be caused both by
enhanced Penning ionization through the excited molecular state and by
radiative decay of the molecule produced. A significant heating is also
observed which can be monitored as a signature of a PA resonance. In this
experiment,  a new cloud has to be trapped, cooled down and detected for
each new choice of the PA laser frequency. Therefore, the PA spectrum can
only be accumulated over small ranges of frequency since the rate of
accumulation of the PA spectrum is of order 1/30 Hz instead of 25 kHz in
the Utrecht experiment. Such a difference in the experimental procedures
comes from the much lower detection efficiency offered by absorption
imaging at 1.083 $\mu$m compared to ionization rate measurement. In
Paris, parts of the spectrum between 0 and 14 GHz below the D2 line have
been recorded. Fig. \ref{Fig:Data_ENS} shows a set of typical spectra.
Since pairs of spin-polarized He* atoms interact through the {\em gerade}
\quintet\ state, only {\em ungerade} molecular states can be excited by
photoassociation.

In both experiments, discrete PA resonances are observed on top of a
broad non-resonant ionization or loss signal which becomes dominant at
high laser intensities and/or small detunings (see
fig.~\ref{Fig:Data_Utrecht} and \ref{Fig:Data_ENS}). Although densities
are \ee{3} times higher and temperatures nearly \ee{3} times lower in the
magnetic trap compared with those in the MOT, we find that most of the
resonance lines in the frequency range of 0-14 GHz below the
\tripS+\tripP\ atomic transition appear in both experiments. The PA lines
are narrower in the Paris experiment because there is no power
broadening, unlike in the Utrecht Experiment. In the present analysis we
will be primarily concerned with discrete line positions. We estimate a
10-20 MHz uncertainty for each line position in both experiments due to
the uncertainty in determining the detuning from atomic resonance using a
Fabry-Perot interferometer. This level of precision is sufficient for a
detailed comparison between the two sets of data and with our calculated
molecular potentials. The energies of the resonances which appear in both
experiments are in full agreement within the experimental uncertainties
\cite{CorrectPreviousMistake}. However, a few PA resonances are observed
in ionization rate detection but not in trap loss. We attribute these to
{\em gerade} excited state potentials not accessible from spin-polarized
atoms, and we will not discuss them in this paper. Conversely, and more
interestingly, a few PA resonances appear only in the trap loss data and
not in the ion data. They bring a new insight in the mechanism of
ionization of these molecules as we shall see below.

\begin{figure}
\twofigures[scale=0.85]{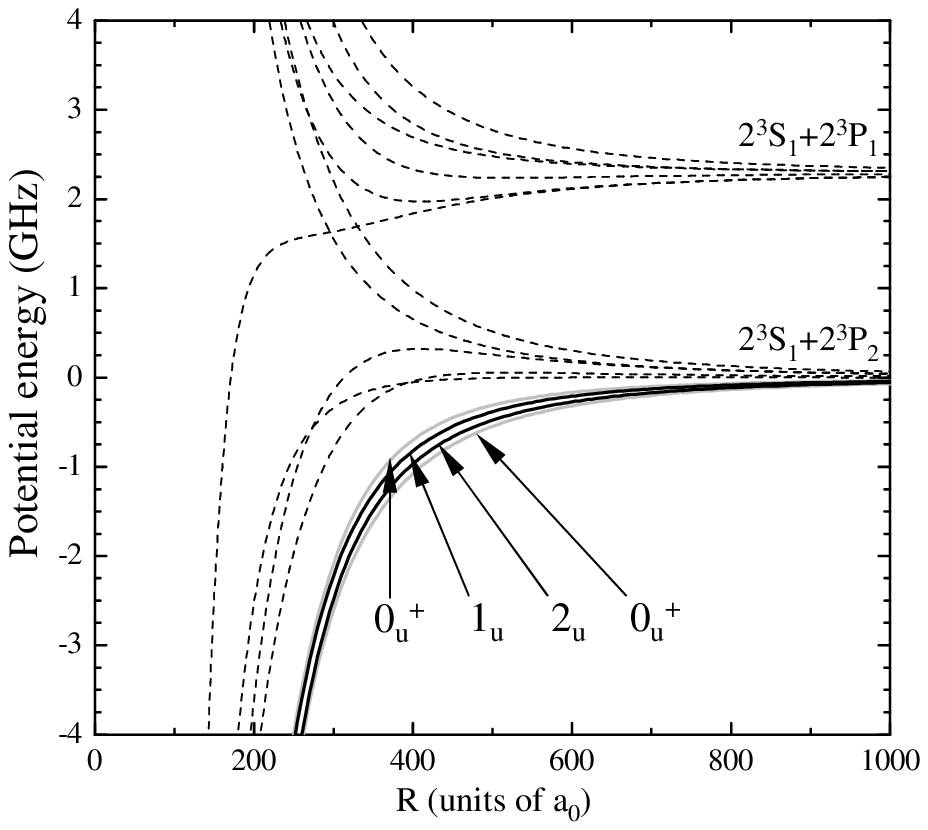}{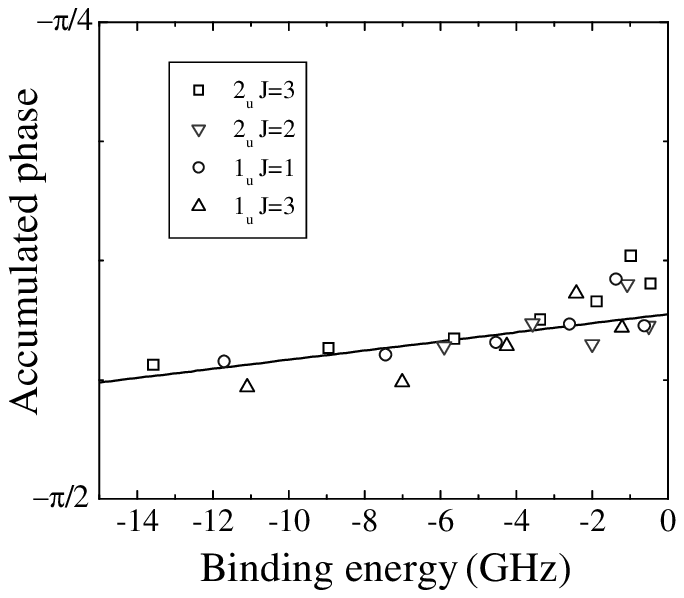}

\caption{{\em Ungerade} Hund's case (c) potentials around the
\tripS-\tripP\ asymptote. The dashed attractive potential curves
correspond to short-range triplet molecular spin states which are
expected to autoionize via the Penning mechanism. The grey solid lines
are the attractive $0_u^+$ states which have partly quintet and partly
singlet spin-character at short interatomic distance. The black solid
lines indicate the $1_u$ and $2_u$ states which become purely quintet at
short range. \label{fg:pot}}

\caption{The accumulated phase of the four series of resonances as a
function of detuning $\Delta\nu$. The resonances correspond to molecular
states which all become $^5\Sigma_u^+$ at short interatomic distance.
\label{fg:AccPhaseResult}}
\end{figure}

We are concerned here by the PA molecular lines corresponding to the {\em
ungerade} excited states near the \tripS+\tripP\ asymptote. Only 27
molecular {\em ungerade} potentials are asymptotically connected to the
pair \tripS+$2^3$P$_{0,1,2}$. The long range tail of these potentials can
be calculated using perturbation theory \cite{Hersbach,PAtheory}. The
resonant retarded dipole-dipole interaction ($C_3/R^3$) and the atomic
fine structure interaction are treated as a perturbation of a pair of
non-relativistic atoms with a fixed internuclear distance $R$. This
approach is clearly valid only for large $R$, where short-range molecular
interactions can be neglected. The analysis shows that some purely
long-range potentials are present near the \tripS+$2^3$P$_{0,1}$
asymptotes, for which it is sufficient to treat molecular rotation and
vibration only as diagonal (first-order) corrections to the electronic
interaction~\cite{ENS2,PAtheory,Venturi}. However, below the
\tripS+\tripP\ asymptote, all the molecular potentials have a short range
part which is not calculated here. The $C_6$ coefficients for the
$\Sigma$ and $\Pi$ electronic states are included in the electronic
interaction, based on the values reported by Venturi
\etal~\cite{Venturi}, but higher-order dispersion terms are neglected.
The potentials of interest for the interpretation of the {\em ungerade}
spectrum are shown in fig.~\ref{fg:pot}. Only 8 {\em ungerade} potentials
have an attractive long-range behavior. Only 4 of these potentials
(displayed in solid line in fig.~\ref{fg:pot}) have significant quintet
spin character at short distance and are expected to give rise to narrow
PA resonances since this stretched state of angular momentum is subject
to the same spin-conservation rule that prevents ionization in a
polarized gas of metastable helium atoms~\cite{Shearer}. For molecular
spin states which are purely triplet or singlet at short interatomic
distance, the ionization probability is so large that such a molecule
would hardly survive half an oscillation period. Consequently, there is a
priori no series of discrete bound states (\idest\ no molecule) to be
expected with strong triplet or singlet spin character.

In order to assign the PA lines to one or several of the attractive, {\em
ungerade} molecular potentials near the \tripS-\tripP, we use the
accumulated phase method~\cite{Verhaar}. Wavefunctions corresponding to a
given electronic state but different (small) binding energies and
different (small) angular momenta should all be in phase at short enough
interatomic distance. In practice, we integrate inwards single-channel
radial Schr\"odinger equations for each of the 8 attractive molecular
potentials and for each of the energies determined experimentally for the
PA resonances:
$$\left\{-\frac{\hbar^2}{m}\;\frac{d^2}{dr^2}+ V_{J,\Omega_{u}}(r)
- E_{res}\right\}\;u(R)=0. $$ Here $m$ is the mass of ${^4}$He, $E_{res}$
is the energy of one of the resonances detected. $V_{J,\Omega_{u}}(r)$ is
the effective long-range interaction potential corresponding to a given
$\Omega_{u}$ {\em ungerade} electronic state and a given rotational state
$J$ when neglecting non-diagonal rotational couplings between $\Omega$
and $\Omega \pm 1$ subspaces \cite{PAtheory,Comment_on_non_diag}. We then
compute the phases $\Phi (r)$ accumulated by the wavefunctions $u(r)$ at
the interatomic distance $r_{in}=20\ a_0$. At this distance, the binding
energy and rotational energy are much smaller than the interaction energy
between the two nuclei and the assumption of stationarity of the
accumulated phase is valid. In addition, the vibrational motion of the
molecule is quasi-classical and the accumulated phases can be written as
$$\Phi (r)\simeq\arctan\left[ \sqrt{m(E_{res}-V_{J,\Omega_{u}}(r))}/(\hbar\,\partial\ln u(r)/\partial r)\right].$$
We plot for each effective potential the accumulated phases as a function
of the PA laser detuning $\Delta \nu$ of the resonances and we search for
series of resonances with nearly equal phases. According to the
assumption above, these series correspond to molecular levels in the same
potential with the same $J$, but increasing vibrational number. This way
we are able to identify 23 resonances in four series as reported in Table
\ref{Tab:_bilan_identification}. The result of the identification does
not depend sensitively on our choice of $r_{in}$ between 15 and 30 $a_0$.
The series found to have a common accumulated phase correspond to the
vibrational progressions of the Hund's case (c) \oneu\ ($J$=1,3) and
\twou ($J$=2,3) states.

\begin{table}[t]

\caption{\footnotesize Measured binding energies for the {\em ungerade}
molecular states observed in the Utrecht and ENS experiments. Four series
of PA resonances are identified which correspond to vibrational levels
belonging to different rotational states in two electronic potentials:
$J=2$ and $J=3$ in $2_u$, and $J=1$ and $J=3$ in $1_u$. The $^*$ signs
indicate resonances which are not detected in Utrecht by ionization-rate
monitoring. All the other lines presented here are observed in both
experiment at the same detuning within the experimental uncertainties of
20 MHz in both cases. Energies are given in GHz with respect to the D$_2$
atomic line.}

\label{Tab:_bilan_identification}
\begin{center}
\begin{tabular}{c|c|c||c|c|c}
\hline
 & $2_u$   &  $2_u$   &  & $1_u$& $1_u$ \\
& $J=2$ &  $J=3$ & & $J=1$ &  $J=3$ \\
\hline \hline
$v$   & -0.51$^*$ & -0.455 & $v'$   & -0.62 & \\
$v-1$ & -1.07$^*$ & -0.98  & $v'-1$ & -1.37 & -1.22\\
$v-2$ & -2.00$^*$ & -1.88  & $v'-2$ & -2.59 & -2.42\\
$v-3$ & -3.57$^*$ & -3.37  & $v'-3$ & -4.53 & -4.25\\
$v-4$ & -5.90$^*$ & -5.64  & $v'-4$ & -7.45 & -7.01\\
$v-5$ &       & -8.95  & $v'-5$ & -11.70& -11.10 \\
$v-6$ &       & -13.56 & $v'-6$ &   &  \\

\hline

\end{tabular}
\end{center}
\end{table}

It turns out that the four series of resonances identified correspond to
molecular states which all become $^5\Sigma_u^+$ with purely quintet spin
character at short range, and hence have the same short-range interaction
potential. Therefore the four corresponding accumulated phases should all
be equal. This is confirmed by plotting on a same graph the accumulated
phases versus detuning for the four series identified as shown in
fig.~\ref{fg:AccPhaseResult}, where all resonances belonging to one
series are indicated with the same symbol. A linear fit of the whole set
of points in fig.~\ref{fg:AccPhaseResult} gives the residual dependence
of the optimal accumulated phase with the energy $E$: $\Phi
(r_{in})\simeq -1.30(2) + 0.0036(20)\times E$. Note that the actual value
of the optimal accumulated phase is a priori wrong since the potentials
we use are only valid at long range and certainly not at distances as
small as $r_{in}=20\ a_0$. However, the only requirement for the method
to work is that the interaction potentials are exact at long interatomic
distances where the calculated wavefunctions are no longer in phase with
each other.

The most important result of the comparison between the two experiments
is that the series of \twou\ ($J$=2) is missing in the ionization data,
whereas the other three series have been detected in both experiments.
Non-diagonal coupling between the $\Omega$ subspaces should be considered
in detail in order to understand this fact. The \oneu\ and \twou\
potentials are close in energy from two \zeroup\ potentials over a large
range of internuclear distances (see fig.~\ref{fg:pot}). The two \zeroup\
potentials connect for a significant part to the \singlet\ potential at
short range, which gives rise to ionization. By contrast, the \oneu\ and
\twou\ become purely quintet, and hence the Penning ionization process is
largely inhibited as already mentioned. However, accurate description
shows that molecular rotation couples \oneu\ to \zeroup\ as well as
\twou\ to \zeroup\ (to the second order, via the nearby \oneu\ state),
allowing for some ionization probability of the \oneu\ and \twou\ states.

Given the properties of the \zeroup\ states with respect to the inversion
($u/g$) and reflection ($\pm$) symmetries, Bose statistics (He$^*$ and
its nuclei are bosons) imposes that the rotational quantum number $J$
must be odd for \zeroup\ states \cite{Hougen}. Therefore the non-diagonal
rotational coupling to a \zeroup\ state can only be effective for odd
values of $J$. Hence, at short distance \oneu\ $J=2$ and \twou\ $J=2$
remain purely quintet whereas \oneu\ $J=1,3$ and \twou\ $J=3$ are
contaminated by non quintet spin states through the coupling to \zeroup\
$J=1$ and $J=3$. This explains why ion detection is only possible for
\oneu\ $J=1,3$ and \twou\ $J=3$ whereas \oneu\ $J=2$ and \twou\ $J=2$
produce no detectable ions in the Utrecht Experiment. The \twou\ $J=2$
resonances appear only in the ENS experiment with an intensity comparable
to that of the other 3 series. The \oneu\ $J=2$ is not detected in the
Paris experiment with a significant signal to noise ratio. However, based
on the above analysis, one resonance observed at -1.27(2) GHz in the
spectrum obtained at ENS could possibly be assigned to the \oneu\ $J=2$
series. Other resonances belonging to the same series are expected at
frequencies which were either not scanned at all at ENS or scanned in a
preliminary experiment with reduced sensitivity, which means that the
intensity of those resonances should be at least 5 to 10 times less than
the other resonances detected. Finally, all the resonances observed in
Paris \cite{EPJDPaperENS} are identified by this analysis except those
which appear at detunings smaller than -0.28 GHz for which the accuracy
of the frequency measurement (20 MHz) does not allow for an accurate
determination of the corresponding accumulated phases. As expected, no
resonance has been observed with a non-quintet spin state at short
interatomic range.

In the Utrecht experiment, the signal-to-noise ratio vanishes when the
detuning is increased and no PA resonance has been observed at detunings
larger than -13.56 GHz. In the ENS experiment, no systematic scan is
possible at large detunings due to the small rate of data accumulation.
Therefore the observation of  resonances detuned further from resonance
is only possible if their position is predicted. The PA line intensities
are modulated by the amplitude of the ground state radial wavefunction
\cite{LineModulation} and one can expect a vanishing line intensity for
excited bound states having their outer turning points $R_{out} \simeq
a$, where $a = 200\pm40\; a_0$ is the s-wave scattering length
\cite{Orsay_a}. This corresponds to PA resonances in the range -5 GHz to
-15 GHz. Deeper bound states (with $R_{out}<a$) should instead lead to
detectable PA resonances. These bound states might prove useful as
intermediate excited states for driving two-photon transitions from a
free pair of atoms to a bound pair in the ground state $^5\Sigma_g^+$.
Indeed, deeper excited bound states have better Franck-Condon overlap
with the final bound state. In addition the background non-resonant
ionization reduces with increasing detuning.

From the present analysis we can conclude, that the ionization and trap
loss experiments yield complementary information on the photoassociation
of the He*-He* system on the \tripS-\tripP\ transition~\cite{Gadea}.  We
have been able to identify four series of resonances and indicated why
those series have a small ionization probability. Of the four series, one
series is only seen in the trap loss experiment. The most striking point
out of this study is, that this \twou\ ($J=2$) potential cannot couple by
rotational coupling to potentials, which have a large ionization
probability. Thus, a molecule in this potential preferentially decays
radiatively. This study might play an important role in future two-photon
PA experiments to be done in Paris for an accurate determination of the
elastic scattering length for the He*-He* system. Indeed, we can use the
potential curves and the accumulated phase determined in the present
analysis, to predict the position of the lower-lying bound states in the
\oneu\ and \twou\ potentials. These states are only accessible
experimentally if one approximately knows where to look for them, given
the low rate of accumulation of the data in the experiment of ENS. They
might prove to have good Franck-Condon overlap with the least-bound state
in the ground-state potential, which is of crucial importance to drive
two-photon transitions in the view of measuring the scattering length of
spin-polarised metastable helium.

The work of the Utrecht group has been supported by the
``Stichting voor Fundamenteel Onderzoek der Materie (FOM)'', which
is financially supported by the ``Nederlandse organisatie voor
Wetenschappelijk Onderzoek (NWO)''. DN is supported by the EU
research training network ``Cold molecules'' (COMOL), under the
contract number HPRN-2002-00290.





\begin{thebibliography}{10}
\bibitem{CorrectPreviousMistake}
The results published in \cite{Hersbach} by the Utrecht group have been
corrected after improving the calibration of the frequency of the PA
laser. The correct measurements are displayed in Table
\ref{Tab:_bilan_identification}.
\bibitem{Comment_on_non_diag}

In the cases where the effective one-channel potential curves
$V_{J,\Omega_{u}^{\pm}}(r)$ don't cross each other the non-diagonal
rotational couplings (between $\Omega$ and $\Omega \pm 1$ subspaces) can
also be accounted for as a correction to these effective potentials.
However, if two effective potential curves cross each other, a full
multichannel calculation is needed. It turns out this situation occurs
only for the effective curve $2_u,J=3$ which crosses $0_u^+,J=3$. In this
case, the non-diagonal coupling is a second order perturbation through
the $1_u$ state and we simply neglect it in our analysis, restricting
ourselves to solving one-channel Schr\"odinger equations.
\end{thebibliography}
\end{document}